\documentstyle[12pt]{article}

\newcommand{\p}{\bot}
\newcommand{\dd}{\partial}
\newcommand{\de}{\delta}
\newcommand{\De}{\Delta}
\newcommand{\om}{\omega}
\newcommand{\Om}{\Omega}
\newcommand{\e}{\varepsilon}
\newcommand{\f}{\varphi}
\newcommand{\ls}{\left(}
\newcommand{\rs}{\right)}
\newcommand{\g}{\gamma}
\newcommand{\al}{\alpha}
\newcommand{\m}{\mu}
\newcommand{\La}{\Lambda}
\newcommand{\ra}{\rangle}
\newcommand{\te}{\theta}
\newcommand{\si}{{\rm sign}}
\newcommand{\str}[1]{\mathrel{\mathop{\longrightarrow}\limits_{#1}}}

\newcommand{\disn}[2]{$$\displaylines{\refstepcounter{equation}%
            \label{#1}\hskip 1em minus 1em #2\hfilneg}$$}
\newcommand{\nom}{\hfil\hskip 1em minus 1em (\theequation)}
\newcommand{\no}{\hfil \hskip 1em minus 1em\phantom{(\theequation)}%
            \hfilneg\cr\hfilneg\hskip 1em minus 1em\hfil}
\newcommand{\ns}{\hfill\cr\hfill}

\makeatletter

\newcount\dobav
\newcount\otlog
\newcount\vrem
\def\@citex[#1]#2{\if@filesw\immediate\write\@auxout{\string\citation{#2}}\fi
  \let\@citea\@empty
  \dobav=-1
  \otlog=-1
  \@cite{\@for\@citeb:=#2\do
    {\def\@tempa##1##2\@nil{\edef\@citeb{\if##1\space##2\else##1##2\fi}}%
     \expandafter\@tempa\@citeb\@nil
     \@ifundefined{b@\@citeb}{\@warning%
       {Citation `\@citeb' on page \thepage \space undefined}%
       \vrem=-1}{\vrem=\csname b@\@citeb\endcsname}
\advance\vrem by -1 \ifnum \vrem=\dobav
 \otlog=\vrem
 \advance\otlog by 1
\else
 \ifnum \vrem=\otlog
  \advance\otlog by 1
 \else
  \ifnum \otlog>0
   \advance\dobav by 1
   \ifnum \otlog=\dobav
    \hbox{,\penalty\@m\ \the\otlog}%
   \else
    \hbox{--\the\otlog}%
   \fi
   \otlog=-1
  \fi
  \dobav=\vrem
  \advance\dobav by 1
  \@citea\def\@citea{,\penalty\@m\ }%
  \ifnum \dobav=-1
   {\reset@font\bf ?}%
  \else
   \hbox{\the\dobav}%
  \fi
 \fi
\fi
}%
\ifnum \otlog>0
 \advance\dobav by 1
 \ifnum \otlog=\dobav
  \hbox{,\penalty\@m\ \the\otlog}%
 \else
  \hbox{--\the\otlog}%
 \fi
\fi }{#1}}

\renewcommand{\section}{\@startsection{section}{1}{0pt}%
          {3.5ex plus 1ex minus .2ex}{2.3ex plus .2ex}{\noindent\hfil\bf}}

\makeatother

\newcommand{\st}{\protect\\ $\protect\vphantom{0}$\hfil}

\begin{document}
\large

\title{The Light-Front Hamiltonian formalism\\
for two-dimensional Quantum Electrodynamics\\
equivalent to the Lorentz-covariant approach\\}

\author{S.~A.~Paston\thanks{E-mail: paston@pobox.spbu.ru},
E.~V.~Prokhvatilov\thanks{E-mail: Evgeni.Prokhvat@pobox.spbu.ru},
V.~A.~Franke\thanks{E-mail: franke@pobox.spbu.ru}\\
St.-Petersburg State University, Russia}

\date{\vskip 15mm}

\maketitle

\begin{abstract}
\normalsize
A light-front Hamiltonian reproducing the results of
two-dimensional quantum electrodynamics in the Lorentz
coordinates is constructed using the bosonization procedure and
an analysis of the bosonic perturbation theory in all orders in
the fermion mass. The resulting Hamiltonian involves a
supplementary counterterm in addition to the usual terms
appearing in the naive light-front quantization. This term is
proportional to a linear combination of zeroth fermion modes
(which are multiplied by a factor compensating the charge and
fermion number). The coefficient of the counterterm has no
ultraviolet divergence, depends on the value of the fermion
condensate in the $\te$-vacuum, and is linear in this value for a
small fermion mass.
 \end{abstract}

\newpage
\section{Introduction}

The Hamiltonian approach to quantum field theory in the
light-front (LF) coordinates $x^{\pm}=(x^0 \pm x^3)/\sqrt{2},\;
x^{\p}=(x^1,x^2)$, where $x^+$ plays the role of time \cite{1},
is one of the nonperturbative approaches to solving the strong
coupling problem \cite{3,4}. In the framework of this approach,
quantization is performed in the plane $x^+=0$, and the generator
$P_+$ of the shift along the $x^+$ axis serves as the
Hamiltonian. The generator $P_-$ of the shift along the $x^-$
axis does not displace the quantization surface and is therefore
kinematic (in the Dirac terminology) in contrast to the dynamic
generator $P_+$. As a result, the momentum operator $P_-$ turns
out to be quadratic with respect to the fields and is independent
of the interaction. On the other hand, this operator is
nonnegative and has a zero eigenvalue only on the physical
vacuum. Therefore, the field Fourier modes corresponding to
positive and negative values of $p_-$ play the role of creation
and annihilation operators over the physical vacuum, and they can
be used to construct the Fock space. In the LF coordinates, the
physical vacuum thus coincides with the "mathematical" vacuum.
The bound-state spectrum can be found by solving the Schrodinger
equation
 \disn{1}{
P_+|\Psi\rangle =p_+|\Psi\rangle
 \nom}
in the subspace with fixed $p_-$ and $p_{\p}$,
which gives the expression $m^2= 2p_+p_- -p_{\p}^2$ for the mass $m$.

This search for bound states can be performed outside the
framework of perturbation theory (PT), for example, using the
so-called discretized LF quantization method (the discretized
light-cone quantization (DLCQ) method \cite{3,5}). But the
LF-Hamiltonian formalism involves a specific divergence at
$p_-=0$ \cite{3,4} and requires its regularization. The simplest
regularization (we refer to it as lightlike) is the ordinary
cutoff $|p_-|\ge\e >0$ breaking the Lorentz and gauge
invariances. It is also possible to apply another regularization
preserving the gauge invariance, namely, the cutoff $|x^-|\le L$
with the introduction of periodic boundary conditions (this
regularization is used in the DLCQ method.) In this case, the
momentum $p_-$ becomes discrete ($p_-=p_n=\pi n/L$, where $n$ is
an integer), and the zeroth field mode corresponding to $n=0$ is
explicitly separated. In principle, the canonical formalism
permits expressing the zeroth mode via the others by solving the
constraints, which is usually a complicated problem \cite{9}.

Introducing the lightlike regularization can generate a
nonequivalence between the LF-coordinate theory and the ordinary
Lorentz-covariant theory. Indeed, experience with nonperturbative
calculations based on the LF-Hamilto\-nian formalism shows that the
calculation result can differ from the corresponding result
obtained in Lorentz coordinates \cite{ann,14,15}. Moreover, such
differences were found even in the lowest PT orders \cite{17}. This
leads to the problem of "improving" the canonical LF Hamiltonian
(resulting from the naive quantization in LF coordinates), i.e.,
to the problem of finding counterterms for this Hamiltonian that
compensate the indicated differences. If this problem can be
solved in all PT orders, then the resulting improved LF
Hamiltonian can be used for nonperturbative calculations.

The PT generated by the LF Hamiltonian can be represented as the
Feynman PT with the same diagrams, but with the cutoff $|p_-|\ge\e>0$
and with a special integration rule, which we call the
lightlike calculation, namely, integration is performed first
with respect to $p_+$ and then with respect to the other momentum
components \cite{har}. To reveal the abovementioned differences in the
PT framework, it therefore suffices to compare the lightlike and
the ordinary (which we call Lorentz) methods for calculating
diagrams. (In what follows, the terms lightlike or Lorentz
applied to diagrams and Green's functions mean that the lightlike
or Lorentz calculation methods are used to find them.)

For nongauge field theories of the type of the Yukawa model, it
is possible to find the counterterms required for improving the
canonical LF Hamiltonian \cite{17,19} including all PT orders in the
coupling constant \cite{20}. But the direct application of the method
in \cite{20} to the gauge theory using the simplest ultraviolet (UV)
regularization requires adding infinitely many counterterms to
the canonical LF Hamiltonian. This difficulty can be overcome by
introducing a specific nonstandard regularization \cite{21} similar to
the Pauli-Villars regularization, and the gauge invariance is then
broken. This leads to the appearance of a large, but finite,
number of counterterms with unknown coefficients. As a result,
the improved LF Hamiltonian for quantum chromodynamics contains
great number of undetermined coefficients, and it reproduces the results of
the Lorentz-covariant theory in all PT orders only for some
preliminarily unknown dependence of these coefficients on the
regularization parameter in the regularization-removal limit. The
presence of unknown coefficients and the complicated structure of
the regularization (the regularized Hamiltonian contains many
additional fields) make practical calculations with the resulting
Hamiltonian extremely complicated. Moreover, because only the PT
with respect to the coupling constant was analyzed, there could
remain purely nonperturbative effects that are not taken into
consideration.

In this paper, a different method is suggested for constructing
an LF Hamiltonian suitable for twodimensional gauge theories (the
two-dimensional quantum electrodynamics (QED-2) is considered
here). We first pass to the boson formulation of the theory and
then analyze the boson PT (with respect to the fermion mass) to
find the improved LF Hamiltonian in terms of bosons. Furthermore,
we return to the fermion variables in the already constructed LF
theory. We note that the boson PT is principally distinct from
the PT with respect to the coupling constant in the original
fermion theory (in QED-2, the latter PT does not exist at all
because of infrared divergences). Therefore, the resulting LF
Hamiltonian can take the nonperturbative effects (with respect to
the ordinary coupling constant) into account.

The suggested method for constructing the LF Hamiltonian can be
applied only to QED-2, but the information obtained in the
analysis of this two-dimensional model can be used to develop new
methods that take the nonperturbative vacuum effects into account
in constructing the LF Hamiltonian for fourdimensional gauge
theories. Such attempts to extract some information about
four-dimensional LF gauge theories from an analysis of the QED-2
appeared recently \cite{mac1}.

\section{Method for constructing the LF Hamiltonian for the QED-2}

The suggested method for constructing the LF Hamiltonian for the
QED-2 is based on the possibility of passing from the QED-2 to an
equivalent scalar theory \cite{24} (of the type of the sine-Gordon
model). This is done using the bosonization procedure, i.e., the
transformation from fermion to boson variables \cite{15,28}. After
this transformation, the mass term of the fermion field in the
QED-2 Hamiltonian becomes an interaction term for the scalar
field, and the fermion mass $M$ becomes a coupling constant in the
boson theory. For $M=0$, the QED-2 is the Schwinger model, and
the boson theory turns out to be a free theory. The PT for the
boson theory (i.e., the PT in $M$) is a chiral PT. The
nontriviality of the quantum vacuum in the QED-2 related to
instantons ($\te$-vacuum) \cite{24,adam} is taken into account
explicitly in the boson theory using the parameter $\te$ in the
interaction term.

The Lagrangian of the boson theory has the form
 \disn{3.0}{
L=L_0+L_I,
 \nom}
 \vskip -2em
 \disn{3.1}{
L_0=\frac{1}{8\pi}\ls\dd_\m\f\dd^\m\f-m^2\f^2\rs,
 \nom}
 \vskip -2em
 \disn{3.2}{
 L_I=\frac{\g}{2}e^{i\te}:e^{i\f}:+\frac{\g}{2}e^{-i\te}:e^{-i\f}:,\qquad
 \g=\frac{Mme^C}{2\pi},\qquad
 m=\frac{e_{\rm el}}{\sqrt\pi},
 \nom}
where $e_{\rm el}$ is an analogue of the electron charge, $C=0.577216$ is
Euler's constant, and the normal-ordering symbol means that the
diagrams with closed lines are excluded from the PT with respect
to $\g$ (this corresponds to the usual meaning of the
normal-ordering symbol in the Hamiltonian).

We consider the Feynman rules for this theory in Lorentz
coordinates. There are vertices of two types with $j$ external
lines ($j=0,1,2,\dots$ here). The factors corresponding to these
vertices are $i^{j+1}e^{i\theta}\gamma/2$
for the first type and $i^{-j+1}e^{-i\theta}\gamma/2$ for
the second type. The vertices without lines ($j=0$) are regarded
as connected subdiagrams. It is convenient to relate part of the
vertex factors $i^{\pm j}$ to the lines that are external relative to the
vertex (i.e., $\pm i$ for each of the lines). The propagator
$\Delta (x) = \langle 0|T(\f (x)\f(0))|0\ra$,
where the free field $\f(x)$ corresponds to the
expansion of Lagrangian (\ref{3.0}), has the form
 \disn{3.4}{
\De(x)=\int d^2k\;e^{ikx}\De(k),\qquad
\De(k)=\frac{i}{\pi}\frac{1}{(k^2-m^2+i0)},
 \nom}
where $d^2k=dk_0dk_1$ and $kx=k_0x^0+k_1x^1$.

On one hand, the resulting PT is simple because it is a
scalar-field theory; on the other hand, it is complicated because
it involves a nonpolynomial interaction. Therefore, there are
infinitely many diagrams in each PT order, and their sum can have
a UV divergence, although each of the diagrams is finite \cite{adam,30}.
It can be easily shown that in the second PT order with
respect to $\g$ (and hence also with respect to $M$), the sum of all
diagrams contributing to the nonvacuum Green's function is UV
finite \cite{30}. This suggests that the situation is the same in the
higher PT orders as well. Indeed, analyzing the PT in the
coordinate space permits proving \cite{hep} that the sum of all
connected Lorentz-covariant diagrams of an arbitrary order $n$ with
respect to $\g$ is UV finite, and there is no divergence for $n>2$
even if the method of adjoining external lines is fixed, whereas
for $n=2$, this is so only after the summation using all such
methods. Only the sum of second-order vacuum diagrams and also
the sum of second-order nonvacuum diagrams with a fixed method of
adjoining external lines remain UV divergent. We note that the
indicated UV finiteness can be proved only for Lorentz-covariant
Green's functions and not for lightlike ones, because some
diagrams are zero in the lightlike calculation, which destroys
the proof. The UV divergences of lightlike diagrams can be
automatically regularized using the lightlike cutoff parameter $\e$.
These divergences appear as $\e\to 0$, and they must be compensated
by counterterms.

The exponentiality of the interaction in the model under
consideration permits reformulating the PT in the language of
superpropagators, i.e., of sums of contributions corresponding to
the versions of connecting a pair of vertices by a different
number of propagators \cite{adam,hep}. The superpropagator is equal to
$e^{\De(x)}$ for a pair of vertices of different type
and to $e^{-\De(x)}$ for
vertices of the same type. In this approach, for a given number
of vertices and a fixed method of adjoining external lines, the
sum of all ordinary diagrams (including the disconnected ones) is
described by a single diagram in which each pair of vertices is
connected by the corresponding superpropagator (the connectedness
is always understood in the usual sense).

The presence of UV divergences in the abovementioned sums of
second-order nonvacuum diagrams requires introducing an
intermediate UV regularization (it is intermediate because the
ultimate values of nonvacuum Green's functions are UV finite and
no renormalization of the Lorentz-covariant theory is needed).
For example, the Pauli-Villars regularization can be taken as the
intermediate regularization. An LF boson Hamiltonian was thus
constructed in \cite{30} (a similar consideration was performed in
\cite{29} for the sine-Gordon model involving no UV divergence). This
Hamiltonian involves a counterterm with a coefficient that is
proportional to the chiral condensate and divergent in the limit
of removing the Pauli-Villars regularization. Therefore, this
regularization should be retained to the end of the calculations.
Because the Hamiltonian remains Pauli-Villars regularized, it is
impossible to return to the former fermion variables and obtain
an LF Hamiltonian in fermion terms characteristic of the original
gauge theory. Therefore, we here use a special UV regularization
for the Lorentz-covariant propagator of the boson field $\f$,
 \disn{6.96}{
\De_{reg}(x)= \cases{\De^{lf}_\e(x), &
$\{|x^-|\le\al\}\cap\{|x^+|\le\al\}$ \cr \De(x), &
$\{|x^-|>\al\}\cup\{|x^+|>\al\}$ \cr},
 \nom}
where
 \disn{4.35}{
\De_{\e}^{lf}(x)=\frac{i}{\pi} \int\limits_{|k_-|\ge\e} dk_-
\int_{-\infty}^{\infty}
dk_+\;\frac{e^{i(k_+x^++k_-x^-)}}{2k_+k_--m^2+i0},
 \nom}
is the lightlike propagator and $\al$ is the UV-regularization
parameter. This regularization has a remarkable property, namely,
after the transformation to LF coordinates, it introduces no
additional modifications in the theory apart from the already
performed cutoff $|p_-|\ge\e >0$ and permits the inverse
transformation to the fermion variables.

Unfortunately, if regularization (\ref{6.96}) is used, then the Lorentz-PT
propagator depends essentially on the parameter $\e$ because the
passage to the limit as $\e\to 0$ corresponds to removing the
regularization. This is why direct comparison of the lightlike
and Lorentz PTs in the momentum space using the method in \cite{20}
becomes impossible because the complicated expression for the
superpropagator in the momentum space does not permit separating
the full dependence on $\e$ for the difference between the
calculation results for the diagrams in the Lorentz and LF
coordinates. Therefore, we must perform the analysis in the
coordinate space.

If the method described in \cite{20} is used, then the contribution of
the domain $p_-\approx \e$, $p_+\approx \frac{1}{\e}$
is taken into account, which
corresponds to the domain of large values of $x^-$,
$x^-\approx \frac{1}{\e}$, in
the coordinate space. Analyzing the convergence of exponential
series in the expansions of superpropagators in this domain leads
to the conclusion that these series can be truncated, i.e., it is
possible to pass to "partial" superpropagators,
 \disn{n1}{
 D_{\pm}^{m}(x)=\sum_{m'=0}^{m}\frac{1}{m'!}\ls
 \pm\De(x)\rs^{m'},\qquad
 e^{\pm\De(x)}=\lim_{m\to\infty} D_{\pm}^m(x).
 \nom}
The results obtained by the method in \cite{20} can then be used. This
procedure is performed in Sec.~4, but it must be preceded by
finding the differences between the lightlike and Lorentz
superpropagators for finite
values of $x^-$ as $\e\to 0$. Usually no such differences between the
propagators appear, but such a contribution in terms of
superpropagators appears in the model under study because of the
"bad" UV behavior of the theory. We must first find precisely
this contribution and compensate it using a counterterm for the
LF Hamiltonian.

\section{Compensating the differences between the lightlike and\st
Lorentz superpropagators for finite values of $x^-$}

The lightlike propagator completely regularized by the condition
$\e\le|k_-|\le V$ can be written in the form \cite{hep}
 \disn{4.36}{
\De_{\e,V}^{lf}(x)= \int\limits_{\e}^V\frac{dk}{k} \;e^{-i\ls
kx^-+\frac{m^2}{2k}x^+\rs\si(x^+)}
 \nom}
(Here, we proceed from expression (\ref{4.35}) with the additional cutoff
$|k_-|\le V$.) On the other hand, the Lorentz propagator can be
brought to a similar form with the momentum cutoff in the Lorentz
coordinates, $|k_1|\le\La$,
 \disn{4.24}{
\De_{\La}(x)=\int\limits_{\e_{\La}}^{V_{\La}}\frac{dk}{k}
\;e^{-i\ls kx^-+\frac{m^2}{2k}x^+\rs\si(x^0)},
 \nom}
where $\e_{\La}\str{\La\to\infty}0$ and
$V_{\La}\str{\La\to\infty}\infty$ \cite{hep}.
For $x^2\ne 0$, the regularization
in expressions (\ref{4.36}) and (\ref{4.24}) can be
removed, after which they coincide. (It should be taken into
account that the sign of the exponent in the integrand function
in (10) becomes inessential after the regularization is removed.)
It hence follows that the related superpropagators also coincide.
For $x^2\approx 0$ and $x^\m\ne 0$,
the behavior of the Lorentz propagator in
the regularization-removal limit is described by the relation
 \disn{4.32}{
\De(x)\sim -\ln\ls-\frac{m^2e^{2C}}{4}(x^2-i0)\rs.
 \nom}
This implies that the Lorentz superpropagator connecting vertices
of different types in the regularization-removal limit behaves as
 \disn{4.34}{
e^{\De(x)}\sim-\frac{4e^{-2C}}{m^2}\frac{1}{x^2-i0}
 \nom}
for the indicated values of $x$. The behavior of the Lorentz
superpropagator $e^{-\De(x)}$ connecting vertices of the same type is
described by the right-hand side of (\ref{4.34}) to the -1st power.

The behavior of lightlike superpropagators in the domain $x^-\approx 0$,
$x^+\ne 0$ coincides with that of the Lorentz superpropagator
because formulas (\ref{4.36}) and (\ref{4.24}) coincide in this domain. In
relation to continuity, the behavior of $e^{-\De^{lf}(x)}$
for $x^+\approx 0$ and $x^-\ne 0$
is the same as that of $e^{-\De(x)}$, and $e^{\De^{lf}(x)}$
behaves like the
distribution ${\cal P}\frac{1}{x^+}$ in the sense of the principal value,
 \disn{4.47}{
 e^{\De^{lf}(x)}\sim -\frac{4e^{-2C}}{m^2}
 \ls{\cal P}\frac{1}{x^+}\rs \frac{1}{2x^--i0\si(x^+)}.
 \nom}
(This can be proved by estimating the integral of $e^{\De^{lf}(x)}$
over a small neighbourhood of the point $x^+=0$ \cite{hep}.)
Summarizing the foregoing, we conclude that
 \disn{4.25d.1}{
e^{-\Delta(x)}=e^{-\Delta^{lf}(x)},
 \nom}
 \disn{4.48}{
e^{\De(x)}-e^{\De^{lf}(x)}= -\frac{2\pi
ie^{-2C}}{m^2|x^-|}\de(x^+).
 \nom}
Relations (\ref{4.25d.1}) and (\ref{4.48})
hold in the sense of distributions on the
class of test functions vanishing for $x^{\m}=0$ and having a support
bounded with respect to $x^-$. If the limit values of the lightlike
superpropagators $e^{\pm\De^{lf}(x)}$ replaced with the regularized
superpropagators $e^{\pm\De_{\e}^{lf}(x)}$ then relations (\ref{4.25d.1})
and (\ref{4.48}) hold as $\e\to 0$
if the size $W$ of the abovementioned support with respect to
$x^-$ satisfies the condition $W\e\str{\e\to 0}0$.

The LF Hamiltonian should now be supplemented with a counterterm
ensuring the improvement of the lightlike
superpropagator $e^{\De_{\e}^{lf}(x)}$
according to formula (\ref{4.48}).
The related counterterm for the action can be taken in the form
 \disn{5.70.2}{
S_c\!=\!\frac{\pi e^{-2C}}{2}\frac{\g^2}{m^2} \int
d^2xd^2y\ls:e^{i\f(x)}e^{-i\f(y)}:-1\rs\times\hfill\no\hfill\times
\de(x^+-y^+)\te(|x^--y^-|-\al)\frac{v(\e(x^--y^-))}{|x^--y^-|},
 \nom}
where the parameter $\al$ (first introduced in formula (\ref{6.96}))
is used
to "cut out" the UV singularity and $v(z)$ is an arbitrary
additionally included rapidly decreasing continuous function
satisfying the conditions $v(0)=1$ and $v^*(z)=v(-z)$.

It can be shown \cite{hep} that such an addition to the action is
equivalent (to within an order-$\e$ correction) to adding the
expression
 \disn{5.69.1}{
-\frac{2\pi i}{m^2}e^{-2C}\de(x^+)
\frac{\te(|x^-|-\al)}{|x^-|}v(\e x^-)
 \nom}
to each lightlike superpropagator $e^{\De_{\e}^{lf}(x)}$.
As $\e,\al\to 0$,
expression (\ref{5.69.1}) coincides
with the right-hand side of (\ref{4.48}) in the
action on functions of the class described after formula (\ref{4.48}).
The indicated equivalence is proved by transforming the
expression for an arbitrary Green's function using
superpropagators. In this case,
the estimate $e^{-\De^{lf}_{\e}(x)}\mid_{x^+=0} = O(\e)$
implied by (\ref{4.36}) is essential.

We note that the presence of the delta function of $x^+$ in (\ref{4.48})
and hence in (\ref{5.70.2}) ensures the locality of the action with respect
to the time $x^+$ and permits passing from the counterterm in (\ref{5.70.2})
for the action to that in the LF Hamiltonian.

\section{Analysis of the lightlike and Lorentz-covariant\st
perturbation theories}

We consider the lightlike PT for Green's functions without vacuum
loops that is generated by the action
 \disn{6.122}{
 S=\int d^2x\ls\frac{1}{8\pi}\ls \dd_\m\f\dd^\m\f-m^2\f^2\rs+
 B:e^{i\f}:+B^*:e^{-i\f}:\rs+\frac{2\pi}{m^2}e^{-2C}|B|^2\times\ns
 \times \int d^2xd^2y\ls:e^{i\f(x)}e^{-i\f(y)}:-1\rs
 \de(x^+-y^+)\te(|x^--y^-|-\al)\frac{v(\e(x^--y^-))}{|x^--y^-|}.
 \nom}
This action differs from that corresponding to Lagrangian (\ref{3.0}) in
the replacement of the coupling constant $\frac{\g}{2}e^{i\te}$ with the
complex parameter $B$ and in the addition of the counterterm
described in the foregoing section. (The coefficient in this
counterterm is related to B in the same way as the coefficient in
(\ref{5.70.2}) is related to $\frac{\g}{2}e^{i\te}$.)

We prove that the lightlike PT generated by action (\ref{6.122}) is
equivalent to the Lorentz-covariant PT for all orders in the
limit as $\e,\al\to 0$. In this case, the expression $B$ depends on
$m$, $\g$, $\te$, $\al$ and $\e$, and it is a power series in $\g$,
 \disn{6.90}{
B=\frac{\g}{2}e^{i\te}+\sum_{k=2}^\infty B_k\g^k.
 \nom}

To analyse this PT, it is convenient to pass from its statement
in terms of the superpropagators $e^{\pm\De(x)}$
to the expression via the
"nonfull" superpropagators $(e^{\pm\De(x)}-1)$.
In the new terms, each pair
of points may or may not be connected by the corresponding
nonfull superpropagator on the condition that only connected
diagrams are considered. This excludes vacuum subdiagrams that
would be taken into account if the full superpropagators $e^{\pm\De(x)}$
were used.

In the ordinary lightlike PT, there is a class of diagrams that
are always zero, namely, the diagrams all of whose external lines
are adjoined to a single vertex. A diagram of this type is called
a "generalized tadpole" (or a GT diagram). From some standpoint,
nonzero GT diagrams do exist in the theory under consideration,
namely, in accordance with the presentation in Sec. 3, the second
term in the right-hand side of formula (\ref{6.122})
for the action can be
replaced by adding expression (\ref{5.69.1}) to each superpropagator
$(e^{\De_{\e}^{lf}(x)}-1)$, after which the
GT diagram may become nonzero. It can be
shown \cite{hep} that only a second-order GT diagram with vertices of
different type, which is simply a product of the two vertex
factors and a superpropagator, becomes nonzero. Only the
additional term (\ref{5.69.1}) contributes to this GT diagram, and this
contribution is equal to  $i|B|^2w$, where
 \disn{6.100.2}{
 w=\frac{2\pi e^{-2C}}{m^2}\int
 dx^-\frac{\te(|x^-|-\e\al)}{|x^-|}v(x^-).
 \nom}
This GT diagram may or may not be adjoined to each of the
vertices. Therefore, its inclusion can be replaced by redefining
the vertex factors, namely, the sums $(B+|B|^2w)$ and $(B^*+|B|^2w)$
can be regarded as the vertex factors instead of $B$ and $B^*$. The
lightlike PT generated by action (\ref{6.122}) thus has the form of a set
of diagrams consisting of nonfull lightlike superpropagators with
the addition of expression (\ref{5.69.1}) (in the case of vertices of
different type). Here, these diagrams contain no GT subdiagrams,
and their vertex factors are $(B+|B|^2w)$ and $(B^*+|B|^2w)$.

With regard to intermediate UV regularization (\ref{6.96}), the GT
diagrams in the Lorentz-covariant PT (generated by Lagrangian
(\ref{3.0})) are nonzero in all orders. Nevertheless, arguing by analogy
with the above, we conclude that their inclusion can be replaced
by redefining the vertex factors, namely, the vertex factors
$\frac{\g}{2}e^{i\te}$ and $\frac{\g}{2}e^{-i\te}$
can be replaced with $A$ and $A^*$, where $A$ is
the sum of all GT diagrams with external lines adjoined to the
vertex $\frac{\g}{2}e^{i\te}$,
 \disn{6.90.1}{
A=\frac{\g}{2}e^{i\te}+\sum_{k=2}^\infty A_k\g^k.
 \nom}
Expression (\ref{6.90.1}) is calculated in the Lorentz coordinates (in all
orders of $\g$ including the first). In view of the results in the
proof of UV finiteness mentioned in Sec.~2, we can conclude that
the expressions $A_k$ are finite for $k>2$ and that $A_2$, is divergent
as $\e\to 0$ or $\al\to 0$ (because $A_2$ is the sum of second-order
diagrams with a fixed method for adjoining external lines). It
turns out here that the divergent part can be separated as \cite{hep}
 \disn{6.86.6}{
A_2=\frac{\g^2}{4}w+const.
 \nom}
The Lorentz-covariant PT regularized according to formula (\ref{6.96}) is
thus a set of diagrams consisting of nonfull Lorentz
superpropagators, these diagrams contain no GT subdiagrams, and
their vertex factors are $A$ and $A^*$.

The lightlike and Lorentz PTs obtained in the above form are
suitable for comparison. We first require that their vertex
factors be equal,
 \disn{6.86.4}{
B+|B|^2w=A.
 \nom}
Furthermore, as shown in Sec.~3, the nonfull lightlike
superpropagators with the addition of (\ref{5.69.1}) (for vertices of
different type) that constitute the lightlike PT coincide in the
limit as $\e\to 0$ in the domain
of finite values of $x^-$ (as $\e\to 0$) with the nonfull Lorentz
superpropagators that constitute the Lorentz PT. As a consequence
of regularization (\ref{6.96}), this also holds for $|x^-|\le\al$.
Therefore,
the difference between the lightlike and Lorentz PTs can appear
only at the expense of differences in the domain of large values
of $x^-$, $x^-\sim \frac{1}{\e}$.
As can be seen from formula (\ref{4.36}), for these
(and greater) values of $x^-$, the lightlike superpropagator
remains finite for any $x^+$ and $\e$. Hence, series (\ref{n1}) for the
lightlike superpropagator can be truncated, and the sum of the
residual series is uniformly small with respect to $x^+$ and $\e$. It
can also be shown that the contribution from the additional terms
(\ref{5.69.1}) is insignificant for large values of $x^-$.
Analytic behavior
(\ref{4.32}) of the Lorentz superpropagator permits moving the
integration contour with respect to $x^+$ away from the singularity
$x^+=0$. Hence, the series can once again be truncated, and the
sum of the residual series is uniformly small with respect to $x^+$
(and there is no dependence on $\e$). This means that the
contribution to the difference between the lightlike and Lorentz
diagrams coming from the domain of large values of $x^-$ does not
change under the passage to "partial" superpropagators (\ref{n1}). This
is equivalent to the passage to finite sums of diagrams in terms
of the ordinary propagators. The comparison method in \cite{20} for
the lightlike and Lorentz methods for calculating diagrams can be
used for these diagrams as well. This method applied to the
scalar theory shows that the differences only exist for GT
diagrams \cite{hep}. But the form of PTs that we consider involves no
GT subdiagrams (see the comments after formulas (\ref{6.100.2})
and (\ref{6.86.6})).

We can thus see that if the parameter $B$ satisfies
condition (\ref{6.86.4}),
then the lightlike PT generated by action (\ref{6.122}) is equivalent to
the Lorentz-covariant PT for all orders in the limit as $\e,\al\to 0$.
Hence, the theory defined by the LF Hamiltonian
 \disn{6.123}{
 H=\int dx^-\ls\frac{1}{8\pi}m^2:\f^2:
 -B:e^{i\f}:-B^*:e^{-i\f}:\rs- 2\pi
 e^{-2C}\frac{|B|^2}{m^2}\times\ns
 \times \int\! dx^-dy^-\! \ls
 :e^{i\f(x^-)}e^{-i\f(y^-)}:-1\rs \te(|x^--y^-|-\al)
 \frac{v(\e(x^--y^-))}{|x^--y^-|}
 \nom}
corresponding to (\ref{6.122}) is perturbatively equivalent in the
regularization-removal limit to the Lorentz-covariant QED-2.

\section{Removing the intermediate UV regularization\st
and returning to the fermion variables}

We find the parameter $B$ from Eq.~(\ref{6.86.4}),
 \disn{6.86.5}{
B=-\frac{1}{2w}+\sqrt{\frac{1}{4w^2}+\frac{A'}{w}-A''^2}+iA'',
 \nom}
where $A'$ and $A''$ are the real and imaginary parts of $A$ and the
sign of the root is chosen in accordance with the lowest order in
expansions (\ref{6.90}) and (\ref{6.90.1}).
As can be seen from formula (\ref{6.100.2}), the
expression $w$ is a function of the product $\e\al$, and it diverges as
$\ln(\e\al)$ as $\e\al\to 0$.

Hamiltonian (\ref{6.123}) and Eq. (\ref{6.86.4})
were derived as a result of
analyzing the PT with respect to $\g$ for a fixed value of the
regularization parameter $\al$. The regularization removal $\al\to 0$ is
therefore impossible in the PT framework. But because the
analysis was performed in all PT orders, we leave the PT
framework in what follows and apply the resulting Hamiltonian in
nonperturbative calculations. We can thus remove the intermediate
UV regularization $\al\to 0$ and hence $w\to\infty$.
Then, proceeding from
the available information about the divergence of $A$ (see formula
(\ref{6.86.6}) and the preceding text), expression (\ref{6.86.5})
can be rewritten as
 \disn{6.86.7}{
B=\sqrt{\frac{\g^2}{4}-A''^2}+iA''=\frac{\g}{2}e^{i\hat\te},
\qquad \sin\hat\te=\frac{2A''}{\g}.
 \nom}
It is interesting that the modulus of the coupling constant turns
out the same as in the original Lorentz-covariant theory. (We
recall that the complex parameter $B$ plays the same role in the
lightlike theory as the expression $\frac{\g}{2}e^{i\te}$ in the
Lorentz-covariant theory.) For large values of $\g$, expression
(\ref{6.86.7})
may involve the root of a negative number, which means that
Eq.~(\ref{6.86.4}) has no solution; hence, the suggested scheme for
constructing an LF Hamiltonian cannot be used for these values of
$\g$.

We also note that because of the special structure of the
counterterm, the zero value of the parameter $\al$ in Hamiltonian
(\ref{6.123}) does not result in the appearance of divergences in matrix
elements.

The scalar theory denned by LF Hamiltonian (\ref{6.123}) can be rewritten
in the form of an LF fermion theory using the inverse
bosonization procedure \cite{hep}. Here, the regularization $|p_-|\ge\e$
is replaced with the cutoff $|x^-|\le L$ with periodic boundary
conditions (as in the DLCQ method mentioned in Sec.~1). In this
case, the zeroth mode with respect to $x^-$ is excluded from
consideration, as a result of which the step $\pi/L$ of the discrete
momentum $p_-$ starts to play the role of the parameter $\e$. To
perform the inverse bosonization procedure, it is convenient to
change the normal ordering of the exponentials in the second term
of Hamiltonian (\ref{6.123}) according to the formula
 \disn{7.128}{
 :e^{i\f(x^-)}e^{-i\f(y^-)}:\;=\;:e^{i\f(x^-)}:\;:e^{-i\f(y^-)}:
 \exp\ls-\sum_{n=1}^\infty\frac{1}{n}e^{-i\frac{\pi}{L}n(x^--y^-)}\rs
 \nom}
and choose a function $v(z)$ of the form
 \disn{7.133}{
 v(z)=|z|\exp\ls\sum_{m=1}^\infty\frac{1}{m}e^{-i\pi mz}\rs
 \sum_{n=-\infty}^\infty\frac{1}{n+\frac{1}{2}}e^{i\pi nz}
 \nom}
(We recall the arbitrariness in the definition of $v(z)$ mentioned
in the comment after (\ref{5.70.2}).) In this case, applying the inverse
bosonization formula \cite{28,15}
 \disn{7.p1}{
 \psi_+(x)=\frac{1}{\sqrt{2L}}e^{-i\om}e^{-i\frac{\pi}{L}x^- Q}
 e^{i\frac{\pi}{2L}x^-}:e^{-i\f(x)}:
 \nom}
permits using fermion variables to write LF Hamiltonian (\ref{6.123})
defining a theory that is perturbatively equivalent to the
Lorentz-covariant QED-2 in the regularization-removal limit (see
\cite{hep}),
 \disn{7.p6}{
 H=\int\limits_{-L}^Ldx^-\biggl(\frac{e_{\rm el}^2}{2}\ls \dd_-^{-1}
 [\psi_+^+\psi_+]\rs^2-\frac{M}{2}\ls R\:e^{i\om}d_0^++h.c.\rs-
 \frac{iM^2}{2}\psi_+^+\dd_-^{-1}\psi_+\biggr),
 \nom}
where
 \disn{7.p6.1}{
R=\frac{e_{\rm el}\:e^C}{2\pi^{3/2}} e^{-i\hat\te(M/e_{\rm el},\te)}.
 \nom}
The field $\psi_+$ satisfies antiperiodic boundary conditions with
respect to $x^-$ and can be expanded with respect to the creation
and annihilation operators as
 \disn{7.p3}{
 \psi_+(x)=\frac{1}{\sqrt{2L}}\ls \sum_{n\ge 1}b_n
 e^{-i\frac{\pi}{L}(n-\frac{1}{2})x^-}+ \sum_{n\ge 0}d_n^+
 e^{i\frac{\pi}{L}(n+\frac{1}{2})x^-}\rs,\no
 \{b_n,b_{n'}^+\}=\{d_n,d_{n'}^+\}=\de_{nn'},\quad
 b_n|0\ra=d_n|0\ra=0.
 \nom}
The expression $Q$ in (\ref{7.p1}) is the charge operator defining the
physical subspace of vectors $|phys\ra$,
 \disn{7.p5}{
 Q=\sum_{n\ge 1}b_n^+b_n-\sum_{n\ge 0}d_n^+d_n,\qquad
 Q |phys\ra=0.
 \nom}
The expression $\om$ in (\ref{7.p6}) is the operator canonically conjugate
to $Q$. The operator $\om$ has the properties \cite{28,15}
 \disn{7.p6.4}{
 e^{i\om}\psi_+(x)e^{-i\om} =e^{i\frac{\pi}{L}x^-}\psi_+(x),\quad
 e^{i\om}|0\ra=b^+_1 |0\ra,\quad e^{-i\om}|0\ra=d^+_0
 |0\ra,
 \nom}
which completely define it, and the square brackets mean that the
zeroth mode with respect to $x^-$ is discarded. The lightlike
momentum operator $P_-$ has the form
 \disn{7.p4}{
 P_- =\sum_{n\ge 1}b_n^+b_n\frac{\pi}{L}(n-\frac{1}{2})+
 \sum_{n\ge 0}d_n^+d_n\frac{\pi}{L}(n+\frac{1}{2}).
 \nom}

It is also interesting that the coefficient $R$ in the ultimate
expression for LF Hamiltonian (\ref{7.p6}) and its defining parameter
$\hat\te$
are related to the values of the vacuum condensates in the
Lorentz coordinates \cite{hep},
 \disn{7.p6.2}{
 {\rm Im}\:R=\langle\Om|:\bar\Psi\g^5\Psi:|\Om\ra,\qquad
 |R|=\frac{e_{\rm el}\:e^C}{2\pi^{3/2}},
 \nom}
 \disn{7.p6.2a}{
 \sin\hat\te =\langle\Om|:\sin(\f+\te):|\Om\ra.
 \nom}
(The second relation in (\ref{7.p6.2}) follows from (\ref{7.p6.1}).)
Here, $|\Om\ra$ is the
physical vacuum, and the normal ordering is performed in the
Lorentz coordinates.

\section{Conclusion}

Using the bosonization procedure and an analysis of the PT in all
orders with respect to the fermion mass $M$, we have constructed LF
Hamiltonian (\ref{7.p6}) defining a fermion theory perturbatively
equivalent to the Lorentz-covariant QED-2 in the continuous limit
as $L\to\infty$. This Hamiltonian involves all terms appearing in the
naive quantization in the LF coordinates. In this case, one of
these terms (the third term in (\ref{7.p6})) does not participate in the
quantization in the Lorentz coordinates and appears as one of the
counterterms restoring the equivalence of the lightlike and
Lorentz PTs. Moreover, the resulting Hamiltonian contains one
more counterterm (the second term in (\ref{7.p6})) proportional to the
zeroth modes of the fermion fields times the phase operator $e^{i\om}$,
which neutralizes their charge and fermion number. The
coefficient of this counterterm is determined by the value of
fermion vacuum condensate (\ref{7.p6.2}) and is linear in this value for a
small fermion mass. Under a certain relation between the
condensate value and the magnitude of the charges $e_{\rm el}$,
Eqs.~(\ref{7.p6.2})
defining the abovementioned coefficient may be incompatible,
which means that the suggested scheme for constructing the LF
Hamiltonian is inapplicable in this case. But if the condensate
ever assumes such values, this can only occur for sufficiently
large fermion masses.

The resulting LF Hamiltonian can be used for nonperturbative
calculations using the DLCQ method. We hope that the information
obtained in the analysis of the given two-dimensional model will
facilitate developing constructive methods for the LF Hamiltonian
that take nonperturbative vacuum effects into account and are
applicable to four-dimensional gauge field theories.

\vskip 1em
{\bf Acknowledgments.} One of the authors (E.~P.) is grateful to
F.~Lenz, M.~Thies, and B.~van~de~Sande for the valuable discussion
and help.

This work was supported in part by the Russian Federation
Ministry of Education (Grant No.~EOO-3.3-316),
the DFG (Grant No.~436 RUS 113/324/0(R),
during E.~P.'s stay at Erlangen University), and a
Gribov "World Federation of Scientists" scholarship.

\end{document}